# Task Assignment in Distributed Systems based on PSO Approach

Mostafa Haghi Kashani *

*Department of Computer Engineering, Shahr-e-Qods Branch, Islamic Azad University, Tehran, Iran*

**Abstract**

In a distributed system, Task Assignment Problem (TAP) is a key factor for obtaining efficiency. TAP illustrates the appropriate allocation of tasks to the processor of each computer. In this problem, the proposed methods up to now try to minimize Makespan and maximizing CPU utilization. Since this problem is NP-complete, many genetic algorithms have been proposed to search optimal solutions from the entire solution space. Disregarding the techniques which can reduce the complexity of optimization, the existing approaches scan the entire solution space. On the other hand, this approach is time-consuming in scheduling which is considered a shortcoming. Therefore, in this paper, a hybrid genetic algorithm has been proposed to overcome this shortcoming. Particle Swarm Optimization (PSO) has been applied as local search in the proposed genetic algorithm in this paper. The results obtained from simulation can prove that, in terms of CPU utilization and Makespan, the proposed approach outperforms the GA-based approach.

## 1. Introduction

Considering the success attained in technology, computer architectures and the created software packets, distributed systems have been used in a wide range of applied softwares. -Task Assignment Problem (TAP) has attracted much attention in recent systems. TAP in distributed systems means assigning a task to the processors of each computer to create efficiency in the system. The propose is also minimizing Makespan (total completion time) and maximizing the CPU utilization. This problem is known as an NP-complete problem [1].

Task Assignment is performed in both dynamic and static methods. In a dynamic method, there is no knowledge about it during performance until he arrives. While in the static method, task Assignment is done before performance time and without change. Therefore tasks must be known beforehand. On the other hand, a task assignment algorithm times a set of tasks with the known communicative and processing features on processors, and this is done for optimizing some efficiency criteria such as Makespan and CPU utilization. In this paper, the focus is on the static method. For solving timing problem different solutions have been presented. The proposed methods are classified into three categories: graph theory-based methods [2], mathematical-based methods [3], [4], and heuristic methods [5], [6]. As referred before, TAP is an NP-complete problem. Therefore applying heuristic techniques is more effective in solving this problem. Heuristic techniques are divided into three types of algorithms: 1. iterative improvement [4], 2. probabilistic optimization, and 3. constructive heuristics. In probabilistic optimization, genetic-based methods have attracted the attention of researchers [7]. In genetic-based methods, the main distinction is made based on the manner of performance and encoding chromosomes for modeling the timing. In optimization, there are two search methods: general and local these algorithms which are based on general optimization are useful if the problem has many minimals in the search environment. In this condition an assigning the correct general minimum is difficult. Among global search methods, a memetic algorithm can be referred to [8], [9]. A memetic algorithm is a population-based approach for heuristic search in genetic problems. Generally, a memetic algorithm is a model of a genetic algorithm in which a separate local search is considered.

---

* Corresponding author. E-mail: mh.kashani@qodsiau.ac.ir

Anyway, in genetic-based methods, the total solution space is searched disregarding techniques which can reduce the complication of optimization. Another shortcoming to be mentioned is that it needs a huge time for timing tasks. In this paper, a new memetic algorithm is proposed to solve the above-mentioned problems, which has been used in local search of Particle Swarm Optimization (PSO) [10], [11].

In Section 2 TAP problem is proposed, and its formulation is stated. In part 3, the applied methods are presented. In part 4, the proposed approach is described. The simulation results are investigated in Section 5, and finally, the conclusion is stated in Section 6.

## 2. Problem formulation

In this paper, TAP is considered with the scenario as follows. In this distributed system, heterogeneous processors and their tasks, after being assigned, are not reversible. Minimizing Makespan and maximizing CPU utilization are the goals to be followed in this problem. Therefore the problem is formulated as follows.

Table 1. Applied symbols in the formulation of problem

| | |
|---|---|
| n | The number of tasks |
| m | The number of processors |
| $r_{uv}$ | The rate of communication delay between $P_u$ and $P_v$ |
| $h_{uv}$ | The required time for transferring one data unit from $P_u$ and $P_v$ |
| $a_{ij}$ | The time of executing $t_i$ on $P_j$ |
| $d_i$ | Data volume related to $t_i$ to be transferred when $t_i$ is executed on a remote processor |
| $f_i$ | Target processor which is selected for $t_i$ to be executed on it |
| $c_i$ | The processor which $t_i$ is worked on it |

- $T = \{t_1, t_2, t_3, ..., t_n\}$ is the set of tasks to be executed.
- $P = \{p_1, p_2, p_3, ..., p_m\}$ is a set of processors in distributed systems. In this model, each processor can only execute one task at each moment. each processor finish the existing task first, and then go to the next task. Also, in the executing time, one task cannot be transferred to another processor.
- R is a $m \times m$ matrix while $r_{uv}$ is in $1 \leq u, v \leq m$ range from matrix R.
- H is a $m \times m$ matrix while $h_{uv}$ is in $1 \leq u, v \leq m$ range from matrix H. $h_{uu} = 0$ and $r_{uu} = 0$.
- A is a $n \times m$ matrix while $a_{ij}$ is in $1 \leq i \leq n$, $1 \leq j \leq m$ range from matrix A.
- D is a vector while $d_i$  $1 \leq i \leq n$ are in D.
- F is a vector while $f_i$  $1 \leq i \leq n$ are in F.
- C is a vector while $c_i$  $1 \leq i \leq n$ are in C.
- The existing load on each processor is the sum of performance time for tasks assigned to that processor. As it is possible for tasks to exist on a processor beforehand, in calculating the existing load on a processor, the performance time for tasks which have been on the processor must be regarded. so.

$$Load(p_i) = \sum_{j=1}^{\substack{No.\,of\,allocated \\ tasks\,on \\ processor\,i}} a_{j,i} + \sum_{k=1}^{\substack{No.\,of\,New\,Assigned \\ tasks\,to \\ processor.i}} a_{k,i} \quad (1)$$

- Makespan timing is defined as the finishing time of the last task on each processor, which equals the maximum load on processors.

$$makespan(T) = \max(Load(p_i)) \quad (2)$$
$$\forall 1 \leq i \leq Number\,of\,processors$$

- The time and cost of task transfer among processors in equation (3) is as follows:

$$CC(T) = \sum_{i=1}^{number\,of\,new\,tasks} \left( r_{c_i,f_i} + h_{c_i,f_i} \times d_i \right) \quad (3)$$

- CPU utilization is obtained by dividing the amount of load that processor(the sum of execution time for a task assigned to the processor) by the total duration of timing. It is illustrated in equation (4) as follows:

$$U(p_i) = Load(p_i) / makespan \quad (4)$$

- The average CPU utilization is obtained by dividing the sum of efficiency of each processor by the number of processors:

$$AveU = (\sum_{i=1}^{No\,of\,processors} U(p_i)) / Number\,Of\,processors \quad (5)$$

- The number of accepted queues for each processor: we should define a threshold for overload or under a load of processors. If the total execution time for the task assigned to each processor is between two extremes of the threshold, the queue will be acceptable, otherwise not. This is calculated obtained from equation (6):

$$AveNoAPQ = NoAPQ / Number\,Of\,processors \quad (6)$$

A queue is regarded for each processor. This queue shows the tasks to be performed by processors.

## 3. Applied methods in the proposed approach

In this section, two methods are described, which have been used as the basis for research in this paper.

### 3.1. Particle Swam Optimization

PSO method was first introduced by Kennedy and Eberhart [13-15] in 1995. PSO method is one of the evolutionary computational algorithms which was inspired by nature and is based on iteration. The inspiration source of this algorithm is the social behavior of animals, like the collective movement of fish, birds and the manner of communicating information among them.

PSO method, similar to many other evolutionary algorithm, like genetic algorithm starts with a primary random population matrix. Unlike genetic algorithm, PSO has no evolutionary operator like crossover and mutation. Each element of the population is called particle, which can be considered as substitute for chromosome in a genetic algorithm. PSO approach is composed of a defined number of particle which acquire primary quality randomly. Two quality is defined for each particle location and speed, which respectively are modeled by a place vector and a speed vector. These particle move in the problem space repeatedly to search the new possible options by calculating efficiency conditions. One variable is allocated for saving the best location of each particle in the past(pbest) and another variable is allocated for saving the best-created location among all particles(gbest). Applying the data obtained from these memories, particles decide how to move in the next stage. In each repetition, all particles move in problem space to find the final optimal point. The efficiency of each particle is evaluated by applying a fitness function the value of which is dependent on optimization problem. Each particle updates a location vector and a speed vector by moving in problem space. In equations (7) and (8), these items have been considered.

$$V_i^{k+1} = wV_i^k + c1 * rand1(\ ) * (pbest_i - S_i^k) + c2 * rand2(\ ) * (gbest - S_i^k) \quad (7)$$

$$S_i^{k+1} = S_i^k + V_i^{k+1} \quad (8)$$

Here w is the inertia factor that is used for controlling the effect of the previous speed on the current speed. $V_i^k$ is the speed of i in k iteration. c1 and c2 are positive constant numbers and rand1 and rand2 are random numbers in range [0, 1]. Speed vector is in range [$-V_{max}$, $V_{max}$].

### 3.2. Genetic algorithm

In the 1970s, the idea of applying a Genetic Algorithm (GA) in optimizations of engineering was proposed by a scientist from Michigan university called John Henry Holland. The main and basic idea of this algorithm is transferring hereditary features by gens. This algorithm is a search technique in computer science to find an approximate solution for optimization and search problem. GA is a specific kind of evolutionary algorithm which applies evolutionary biological techniques like mutation and inheritance [16], [17], [18].

Before executing GA for a problem, a method must be used for encoding genomes in computer language. One of the ordinary methods of encoding is binary strip another similar method of encoding solutions is an array of whole numbers and decimal numbers and each state illustrates one aspect of the features. Compared to the previous one, this solution is more complex. GA method searches on a population of solutions, not just on a single point which is a distinguishing feature. This algorithm compared to other conventional optimization methods. Each iteration in GA is composed of a competitive selection that deletes weak solutions from the population. Appropriate solutions along with other solutions, are created by the replacement of a part of a solution with the other which is called crossover. After selection and crossover, there will be a new population of all samples, some of them will be copied directly and the others will be produced by crossover. For the samples not to be repetitive, a little change is created by mutation. A mutation is done by a little change in the solution. Crossover and mutation are used for producing new solutions from appropriate solutions observed in the problem space. A mutation operator is required to make sure of genetic diversity in a population. Pseudo-code related to GA action is as follows:

```
1  Genetic Algorithm
2  begin
3    choose initial population
4    repeat
5      Evaluate the individual fitnesses of a certain proportion of the population
6      Select pairs of best-ranking individuals to reproduce
7      Apply crossover operator
8      Apply mutation operator
9    until terminating condition
10 end
```

Memetic algorithms are a form of GA which provides a local search process for reaching better solutions for difficult problems [9], [19], [20]. In this paper, PSO has been used as a local search in a memetic algorithm.

## 4. The proposed method

In the proposed model, there are a limited number of tasks in the task pool. Each task has a number and a time for execution. Tasks are allocated to processors from that pool. The proposed chromosome is illustrated in Fig. 1. In this chromosome tasks 1, 2, 3, 4, 5, and 6 from left to right have been allocated respectively to processors 4, 1, 4, 2, 2, and 3. Tasks 1 and 3 have been allocated to processor 4, but task 1 is executed and then task 3.

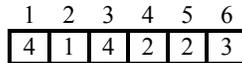

Fig. 1. An example of a proposed chromosome

The fitness of chromosome T can be computed in equation (9)

$$fitness(T) = \frac{(\gamma \times AveU) \times (\theta \times AveNoAPQ)}{(\alpha \times Makespan(T)) \times (\beta \times CC(T))} \quad (9)$$

In this relation $0 < \alpha, \beta, \gamma, \theta \leq 1$ are controlling parameters that are regulated according to special items for the effect of each section. Here the default value is regulated as one. This relation shows that an appropriate solution has a smaller Makespan and less communication cost, and also it has more processor utilization and more acceptable average number of processor queues.

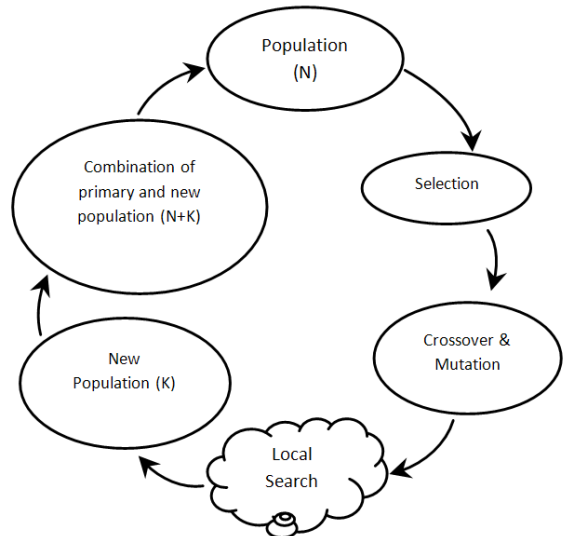

Fig. 2. The structure of the proposed approach

The proposed approach is described in detail. Figure 2 shows the proposed method. PSO is used as a local search in the proposed memetic algorithm. The proposed approach is used as explained below:

The most important case in applying the PSO method is its concordance with the problem. Therefore, all tasks are arranged based on their time of performance. The location of a particle can be produced randomly based on the time order of tasks on different processors. The location of the particle corresponds with the location of the task on each chromosome. In this case, continuous location convert of particles changes into a permutation of tasks. The speed of each particle is calculated by equation (7).

Meanwhile, the value of each particle in $S_i$ in relation (7) can only rest in [-$S_{max}$, $S_{max}$] range which, in itself, leads to the excessive movement of particles out of search space. $S_{max}$ is the maximum time for performing all tasks. The speed of task i in iteration k+1 is calculated based on the best local executing time and the best general executing time in k iteration of itself. The average executing time of task i in repetition k+1 is calculated by equation (8).

## 5. Experimental results

The approach of Particle Swarm Optimization applied in the memetic algorithm has been encoded in Matlab software and in a personal computer with 2GB RAM, Windows XP. In this section, parameters of CPU utilization and Makespan are evaluated in the proposed approach with [21].

The first experiment:

In this experiment, the number of tasks was increased, and the average CPU utilization and Makespan were calculated. The simulation results are illustrated in Figures 3 and 4. As shown in Fig. 3, in the proposed approach, Makespan is less than that of the genetic-based approach.

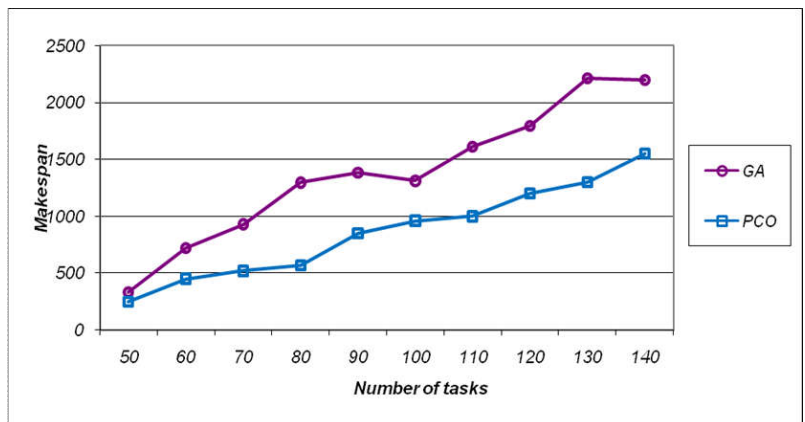

Fig. 3. Makespan in both methods considering the number of tasks

As shown in Fig. 4, the average CPU utilization in the genetic-based approach is less than that of the proposed method.

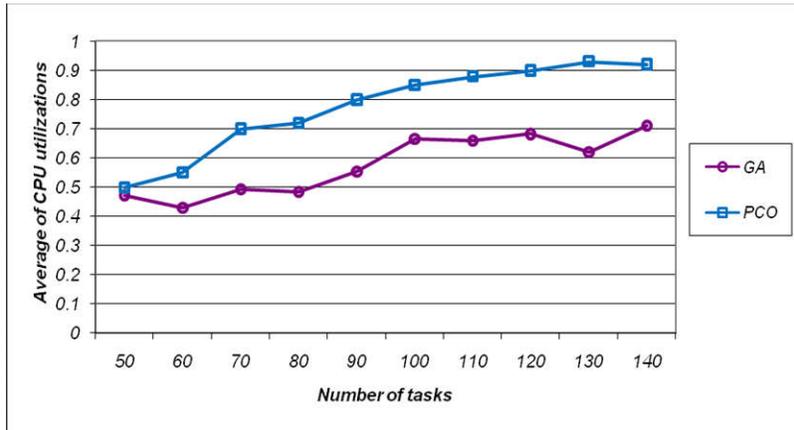

Fig. 4. CPU utilization in both methods considering the number of tasks

The second experiment:

In the second experiment, the aim is to the calculation of the average CPU utilization and Makespan on all processors if the population increases. The simulation results are shown in Figures 5 and 6. If population increases, as illustrated in Fig. 5, Makespan in the genetic-based approach is more than that of the proposed approach. As observed in the figures, the genetic-based method is not scalable.

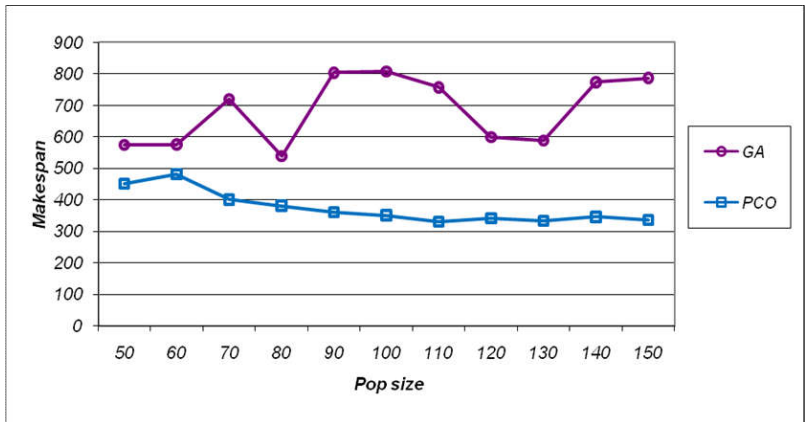

Fig. 5. Makespan in both methods considering population size

Figure 6 illustrates the average CPU utilization in the proposed approach is more than that of the genetic-based method.

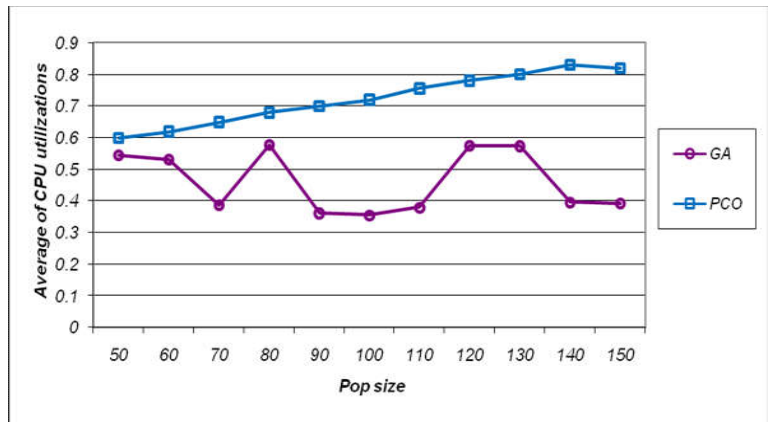

Fig. 6. CPU utilization in both methods considering population size

The third experiment:
In this experiment, the number of generations is increased, and the results are illustrated in Figures 7 and 8.

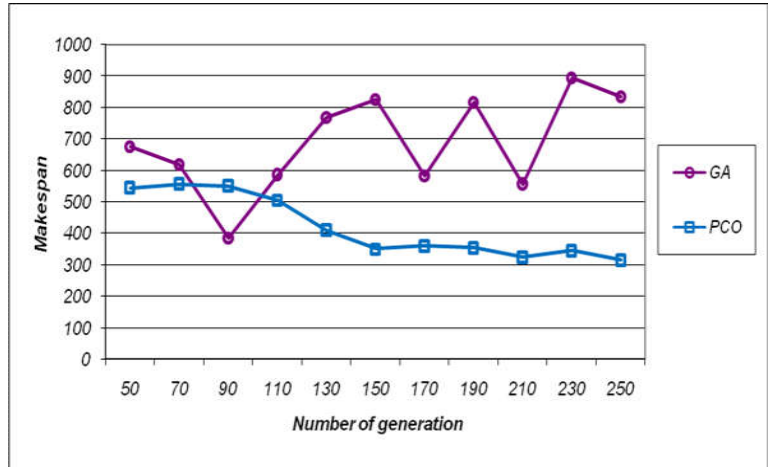

Fig. 7. Makespan in both methods considering the number of generations

In Fig. 8, it can be observed that the average CPU utilization in the genetic-based method is less than that of the proposed approach.

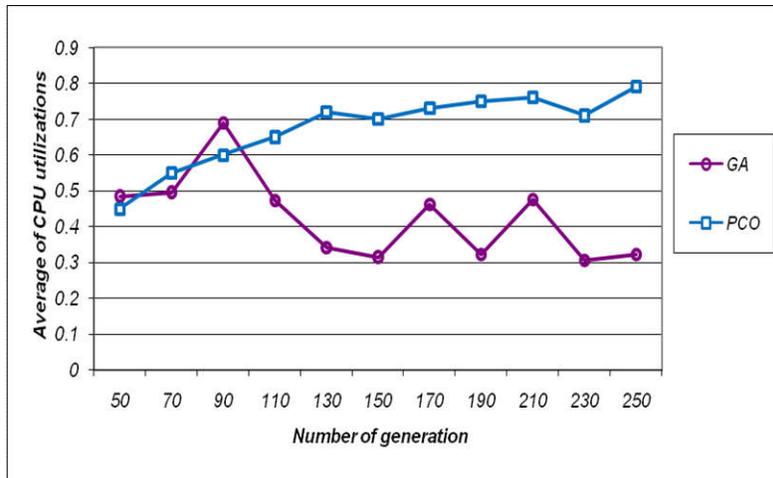

Fig. 8. CPU utilization in both methods considering the number of generations

## 6. Conclusion

Assigning tasks in distributed systems play an important role in the power and efficiency of the whole system. TAP problem, in the best condition, is regarded as NP-complete. In this paper, the approach of Particle Swarm Optimization in local search of the memetic algorithm has been applied, and the proposed approach has been applied for solving the TAP problem. This algorithm follows some goals in evaluating the proposed solution and has solved the TAP problem by minimizing Makespan and maximizing CPU utilization. Most existing approaches just focus on one of these aims. The simulation results prove that the proposed approach has a better performance on Makespan and CPU utilization parameters compared to the genetic-based method.

## 7. Acknowledgments

The researcher is grateful to the research unit of the Islamic Azad University, Shahr-e-Qods branch, for providing the opportunity to conduct this research.